\documentclass[twoside]{article}
\usepackage{fleqn,espcrc2}
\usepackage{epsfig}

\title{Supersymmetry and Strongly Coupled Gauge Theories}

\author{Philip C. Argyres\address{Newman Laboratory, 
Cornell University, Ithaca NY 14853, U.S.A.}\thanks{Supported 
in part by NSF grant PHY95-13717; Cornell preprint CLNS 01/1722.}}
       
\begin{document}

\begin{abstract}
I briefly review how supersymmetry helps in the extraction of exact
nonperturbative information from field theories, and then discuss
some open problems in strongly coupled gauge theories.
(Talk given at ``30 Years of Supersymmetry'' symposium in Minneapolis, 
Minnesota on October 15, 2000.)
\vspace{1pc}
\end{abstract}

\maketitle

Unlike many of the other talks at this symposium, this one will not
take supersymmetry ``seriously'': instead of being concerned with how
or if supersymmetry enters physics beyond the Standard Model, this
talk will treat supersymmetry merely as a technical tool to shed light
on the strong coupling behavior of gauge field theories in general.
Supersymmetric field theories are especially simple and symmetrical
field theories, and as such are particularly ``well behaved''.  We
study the unusual or singular behavior of supersymmetric theories with
the expectation that whatever can happen in a supersymmetric vacuum,
will happen in spades in non-supersymmetric theories.

In the first part of this talk I will briefly outline those features
special to supersymmetric theories which together provide a powerful
handle on the nonperturbative vacuum structure of these theories.  The
second part will then describe some of the more unusual or unexpected
things we have learned about strongly coupled gauge theories.  This
discussion will be formulated in the guise of three open questions.
Here I restrict myself to Lorentz invariant gauge field theories
without gravity in 4, 5, and 6 dimensions with 8 or 16 hermitian
conserved supercharges.  These restrictions are just to keep this talk
short: field theories in lower dimensions and with fewer
supersymmetries have an even greater variety of strong coupling
behaviors, and likewise for gravitational ({\em i.e.}\ string)
theories.

\section{HOW SUPERSYMMETRY HELPS}

Supersymmetric nonrenormalization (NR) theorems aid in the
identification of exactly marginal operators, and ensure exact flat
directions giving rise to moduli spaces of vacua.  The structure of
superconformal algebras aids in the determination of exact anomalous
scaling dimensions.  And the structure of supersymmetry algebras with
central charges can determine the exact mass spectrum of states
invariant under a fraction of the supersymmetry generators (the BPS
states).  We will discuss each of these topics in turn.

\subsection{Exactly marginal operators}

Knowledge of exactly marginal operators (and their associated
dimensionless couplings) is crucial to probing the strong coupling
physics of field theories since one way of approaching strong coupling
is by appropriately tuning a marginal coupling to a large value.  The
NR theorems of supersymmetric field theories give information to all
orders in perturbation theory (see the talks by Iliopoulos and West)
and nonperturbatively (following the work of Seiberg \cite{s9309}) about
certain classes of couplings.

To take the most famous example, in four-dimensional ($d=4$) $N=1$
supersymmetric gauge theories, the superpotential and gauge kinetic
terms,
\begin{equation}
S \supset \int d^4x \int d^4\theta \left\{ \lambda {\cal O}(\Phi)
+ {1\over g^2} W_\alpha^2 \right\} ,
\end{equation}
are protected by such an NR theorem.  More precisely, the theorem says
that the anomalous dimensions $\gamma_i$ of the chiral superfields
$\Phi_i$ (coming from their kinetic terms) can only enter into the
renormalization group (RG) running of the couplings $\lambda$ and
$g^2$ in a prescribed manner.  Suppose, for instance, that
\begin{equation}
{\cal O} \equiv \prod_i \Phi_i^{n_i} .
\end{equation}
Then the NR theorems imply the exact RG $\beta$ function relations
\cite{nsvz83}
\begin{eqnarray}\label{betafncs}
\beta_g &\propto& {3\over2} T({\bf adj}) - {1\over2} \sum_i T({\bf R}_i)
(1-\gamma_i),\nonumber\\
\beta_\lambda &\propto& 3 - d_{\cal O} - {1\over2} \sum_i n_i \gamma_i,
\end{eqnarray}
where ${\bf R}_i$ is the gauge group representation of $\Phi_i$,
$T({\bf R}_i)$ is its index, and $d_{\cal O}$ is the canonical scaling
dimension of ${\cal O}$.

This NR theorem helps in finding exactly marginal operators in those
cases where the choice of representations ${\bf R}_i$ and operators
${\cal O}$ are such that $\beta_g \propto \beta_\lambda$.  In this
case one can deduce the existence of linear combinations of couplings
with exactly vanishing $\beta$ functions (see \cite{ls9503} and references
therein).

A simple example \cite{ls9503} illustrating this is the $d=4$ $N=1$
$SU(n)$ theory with three adjoint chiral multiplets $\Phi_1$,
$\Phi_2$, and $\Phi_3$, with superpotential
\begin{eqnarray}
{\cal W} &=& {\rm tr}\bigl[
a \Phi_1\Phi_2\Phi_3 + b \Phi_3\Phi_2\Phi_1\nonumber\\
&& \ \ {} + c (\Phi_1^3 + \Phi_2^3 + \Phi_3^3) \bigr] .
\end{eqnarray}
{}From the symmetry under cyclic permutations of the $\Phi_i$ we see
that the $\gamma_i$ are all equal, to $\gamma$, say.  Then
(\ref{betafncs}) implies $\beta_g \propto \beta_a \propto \beta_b
\propto \beta_c \propto \gamma$.  Thus, in this four complex
dimensional space of couplings $\{g,a,b,c\}$ we would generically
expect to find a three dimensional submanifold of exactly marginal
couplings; furthermore, since this submanifold goes through the origin
({\em i.e.}\ weak coupling) we are assured that it indeed exists.  The
precise equation for the manifold of marginal couplings
$\gamma(a,b,c,g)=0$ is in general not known exactly away from the
origin.  (In this particular example, though, a one dimensional
subspace is, namely $\{c=0, a=-b=g\}$ where there is an enhanced $N=4$
supersymmetry.)

\subsection{Moduli spaces of vacua}

Another way of probing the strong coupling behavior of field theories
is by tuning the vacuum expectation values (vevs) of scalars to
approach a strongly coupled vacuum.  This can only be done if the
scalars are {\em moduli} of the theory---{\em i.e.}\ have exactly flat
potentials.

The nonperturbative control over potential terms implied by the NR
theorem discussed above allows us in the case of 4 conserved
supercharges ({\em e.g.}\ $d=4$ $N=1$ supersymmetry) to tune the
couplings to ensure the existence of flat directions.  In the case of
8 or more supercharges ({\em e.g.}\ $d=4$ $N\ge2$ or $d=6$
$N\ge(1,0)$) the flat directions are generic---no tuning is necessary.

A well known example is the $d=4$ $N=2$ $SU(n)$ superQCD theory.  Here
the complex scalar $\phi$ in the vector multiplet has an $(n-1)$
complex dimensional moduli space, the Coulomb branch, where
$\langle\phi\rangle$ generically breaks $SU(n) \to U(1)^{n-1}$.  In
the cases where the superQCD theory is asymptotically free (AF) with
strong coupling scale $\Lambda$, the Coulomb branch breaking for
$\langle\phi\rangle \gg \Lambda$ is just a weakly coupled Higgs
mechanism.  On the other hand, sending $\langle\phi\rangle \to 0$
probes strongly coupled $SU(n)$ physics.  As pioneered in
\cite{sw9407,sw9408}, using analytic continuation in $\phi$ together with
some qualitative physical assumptions allows an exact determination of
the nonperturbative low energy effective action on the Coulomb branch
for all values of $\phi$.

More generally, this program of analytic continuation on a moduli
space of vacua is greatly aided by the fact that the geometry of the
moduli space is constrained by supersymmetric selection rules.  For
example, as discussed in the talks of De Wit and Ferrara, $d=4$ $N=1$
supersymmetry implies that the moduli space is a Kahler manifold;
$d=4$ $N=2$ implies a kind of product of rigid special Kahler and
hyperKahler manifolds; while $d=4$ $N=4$ theories have locally flat
moduli spaces with orbifold singularities.

Finally, it is sometimes useful to note that the existence of exactly
marginal couplings and exactly flat directions are not unrelated.  By
embedding a given theory, $A$, in a larger AF theory, $B$, marginal
couplings in $A$ can often be realized as vevs along a submanifold of
the moduli space of $B$.  Indeed, in gravitational (string or M)
theories there are no couplings: all field theory marginal couplings
are lifted to flat directions for some scalars.

\subsection{Superconformal algebras}

The structure of superconformal algebras (discussed in the talks by
Ferrara and West) allows the determination of the exact anomalous
scaling dimensions of certain operators, which often provide useful
``diagnostics'' for strongly coupled vacua.

If a scale invariant supersymmetric vacuum is also conformally
invariant, then the theory is invariant under the superconformal
extension of the supersymmetry algebra.  Thus, heuristically, if the
supersymmetry algebra is
\begin{equation}\label{QQ}
\{ Q_n, Q_m \} = \delta_{nm} \not\!\! P
\end{equation}
where $n$, $m$ run over the different supersymmetry generators in
extended supersymmetry, then the superconformal algebra has double the
number of fermionic generators, with a superconformal generator $S_n$
for each $Q_n$, satisfying
\begin{equation}\label{SS}
\{ S_n, S_m \} = \delta_{nm} \not\!\! K
\end{equation}
where $K_\mu$ is the generator of special conformal transformations.
Associativity of the algebra then implies
\begin{equation}\label{QS}
\{ Q_n, S_m \} = \delta_{nm} D + R_{nm} + \cdots
\end{equation}
where $D$ is the generator of dilatations and $R_{nm}$ are the
R-symmetry generators.  These algebras are tightly constrained by
associativity; for example, for $d>4$ they only exist for $N=1$ in
$d=5$ and $N=(1,0)$ or $(2,0)$ in $d=6$, and not at all if $d\ge7$
\cite{n78}.

The classification of unitary (or ``positive energy'') representations
of the superconformal algebras \cite{m77,k77,dp85} imply
inequalities on the eigenvalues of $D$ and $R_{nm}$ following
essentially from the positivity of operators like $[(Q_n+Q_n^\dagger)
+(S_n+S_n^\dagger)]^2$ and the algebra (\ref{QQ}--\ref{QS}).  These
inequalities are saturated for certain chiral fields (annihilated by
some of the $Q_n$ and $S_n$ generators).  Knowledge of the R-symmetry
charges of these operators then allows one to deduce their $D$
charges, {\em i.e.}\ their exact scaling dimensions \cite{fiz74,cw93}.

\subsection{BPS states}

BPS states are supersymmetric particle states whose mass can be
determined exactly from the structure of the supersymmetry algebra.
They can thus provide another ``diagnostic'' of (non-scale-invariant)
strongly coupled vacua.

The argument is analogous to that determining scaling dimensions from
the superconformal algebra.  In the non-conformal case the R-charges
do not enter the algebra, but a set of $U(1)$ charges---the central
charges $Z_{nm}$---commuting with all other generators can enter:
\begin{equation}\label{QQZ}
\{ Q_n, Q_m \} = \delta_{nm} \not\!\! P + Z_{nm} .
\end{equation}
Positivity of operators like $(Q_n+Q_n^\dagger)^2$ then give
inequalities between the eigenvalues of $P_\mu$ and $Z_{mn}$, {\em
i.e.}\ between particle masses and their $U(1)$ charges.  For certain
states---the BPS states---which are annihilated by some fraction of
the $Q_n$, these inequalities are saturated and the mass is determined
by the central charge \cite{wo78}.  In certain cases the
supersymmetric selection rules and analytic continuation techniques
mentioned in Section 1.2 above can be used to determine the central
charges and thus the BPS spectrum as functions of the couplings and
vevs at strong coupling; {\em e.g.}\ \cite{sw9407,sw9408}.

\section{THREE QUESTIONS}

Instead of attempting a survey of the kinds of strongly coupled
phenomena that have been found using the tools outlined in the last
section, I will focus on three open questions concerning strongly
coupled phenomena which I think are both important and on which
progress can be made.

\subsection{``Ultra-strong'' coupling}

Different values of the exactly marginal complex gauge coupling
\begin{equation}
\tau \equiv {\theta\over2\pi} + i {4\pi\over g^2},
\end{equation}
where $g^2$ is the gauge coupling and $\theta$ the theta angle in a
scale invariant $d=4$ gauge theory, are typically identified under
S-duality (also known as Montonen-Olive duality or strong-weak
coupling duality) transformations.  The prototypical example of this
\cite{mo77,gno77,wo78,o79,s9402} is the $SL(2,{\bf Z})$ group of
S-duality identifications in $d=4$ $N=4$ superYang-Mills.  This group
is generated by the theta angle $2\pi$ rotation $\tau\to\tau+1$ and by
inversion of the gauge coupling $\tau\to-1/\tau$.  This S-duality
group relates weak ($g\to0$) to ultra-strong ($g\to\infty$) couplings,
but does not ``solve'' the strong coupling behavior of $N=4$
superYang-Mills.  This is illustrated in Figure~1 which shows a
fundamental domain in the complex $\tau$ upper half plane under these
identifications.  This domain includes a weak coupling limit ($g\to0$)
at $\tau=+i\infty$ as well as some special strong coupling vacua at
$\tau=i$ and $e^{i\pi/3}$ (where $g \sim 1$).  These special strong
coupling points have enhanced ${\bf Z}_2$ and ${\bf Z}_3$ discrete
global symmetries, respectively, but apparently no other more striking
behavior.  A similar story holds for scale invariant $d=4$ $N=2$
$SU(2)$ superQCD with four fundamental hypermultiplets
\cite{sw9407,sw9408}.

\begin{figure}[htb]
\centerline{\psfig{figure=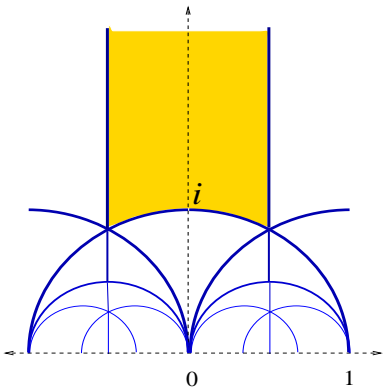,width=2in,height=2in}}
\caption{The shaded region is a fundamental domain of the action of
$SL(2,{\bf Z})$ in the complex $\tau$ plane.}
\end{figure}

However, the story changes qualitatively for $d=4$ $N=2$ superQCD with
gauge groups with rank $\ge 2$.  For the generic such theory, the
S-duality relations no longer identify all ultra-strong coupling
points with weak coupling points.  This is illustrated in Figure~2 which
shows a fundamental domain for the S-duality group of $SU(n)$ $n\ge3$
$d=4$ $N=2$ superQCD \cite{aps9505} which is an index three subgroup
of $SL(2,{\bf Z})$.  Note that a point ($\tau=0$) with $g=\infty$ is
included in the domain.

\begin{figure}[htb]
\centerline{\psfig{figure=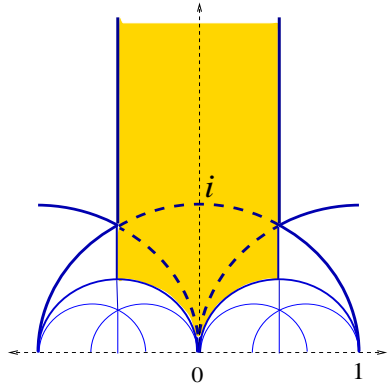,width=2in,height=2in}}
\caption{The shaded region is a fundamental domain of the action of
an index 3 subgroup of $SL(2,{\bf Z})$ in the complex $\tau$ plane.}
\end{figure}

The question thus arises as to what is the nature of the physics at
these ultra-strongly coupled theories with $\tau=0$.  That there is
some qualitatively new physics (as compared to the merely strongly
coupled theories) can be deduced from the behavior of the low energy
$U(1)^{n-1}$ effective action on the Coulomb branch of these theories,
where we observe that as $\tau\to0$ the effective action becomes
singular, indicating the appearance of new light degrees of freedom.
Moreover, these new singularities appear everywhere on the Coulomb
branch and not just on a submanifold of it.

There are a number of possibilities for the answer to this question.
Perhaps these ultra-strong coupling points are in fact the weak
coupling limits of some other theory.  If so this would raise the
possibility of a ``web of dualities'' connecting different field
theories similar to what is seen in Calabi-Yau compactifications of
string theories.  However in the simplest case of $SU(3)$ $N=2$
superQCD an explicit survey of all known scale-invariant theories with
rank 2 gauge groups finds no weak coupling candidate which matches
onto the $SU(3)$ theory.  For example, a promising candidate (from
hints from a IIA string construction of the $SU(3)$ theory) is the
scale invariant $SU(2)\times SU(2)$ theory with 2 fundamental
hypermultiplets in each factor and one bifundamental hypermultiplet;
an analysis \cite{ab9910} of the coupling space of this theory shows
no ultra-strong coupling points.

A second possibility is that due to an ambiguity in the overall
scaling of operators as one approaches the ultra-strong point, the
singularity in the effective action is really just a kind of
coordinate singularity---{\em i.e.}\ a place where no new light states
occur, but where the coupling space joins smoothly to that of another
theory.

A final, and most likely, possibility is that the ultra-strong
coupling points signal a genuinely new strongly coupled ``phase'' of
gauge theories.  Perhaps a hint of this comes from the IIA brane
construction \cite{w9703} for the $N=2$ theories, where the
ultra-strong points correspond to brane configurations with coincident
NS5 branes.  We will discuss the $d=6$ physics of stacks of NS5 branes
in Section~2.3 below.

\subsection{Gauge theories with no weak coupling limits}

Another interesting class of strongly coupled theories occur as
superconformal vacua at special points on a moduli space of vacua.
For example, all the $d=4$ $N=2$ superconformal field theories with a
one complex dimensional Coulomb branch (which we will parameterize as
the complex $u$ plane) can be described by the singular degenerations
of Seiberg-Witten tori as shown in Table~1.

\begin{table}[htb]
\caption{The physical inequivalent degenerations of Seiberg-Witten
tori (as $u\to0$) describing $d=4$ $N=2$ conformal field theories with
a single modulus $u$ together with their associated global symmetry
groups $G$ and the scaling dimensions $D(u)$ of the modulus superfield.}
\begin{tabular}{@{}lllll}
\hline
Curve			& $G$ 		& $D(u)$\\
\hline
$y^2 = x^3+u$		& $U(1)$	& $6/5$ \\
$y^2 = x^3+ux$		& $U(2)$	& $4/3$ \\
$y^2 = x^3+u^2$		& $U(3)$	& $3/2$ \\
$y^2 = x^3+\lambda u^2x+u^3$ & $SO(8)$ 	& $2$ \\
$y^2 = x^3+u^4$		& $E_6$		& $3$ \\
$y^2 = x^3+u^3x$	& $E_7$		& $4$ \\
$y^2 = x^3+u^5$		& $E_8$		& $6$ \\
\hline
\end{tabular}
\end{table}

Here $x$ and $y$ are the auxiliary complex variables describing the
Seiberg-Witten Riemann surface.  The mass scaling dimensions $D(u)$ of
the $N=2$ chiral superfield $U$ whose lowest component is the modulus
$u$ are found using the superconformal algebra as described in
Section~1.3 above.  The dimensions of the associated coupling
$\lambda$ is $D(\lambda) = 2 - D(u)$ as follows from the form of the
$N=2$ superspace coupling
\begin{equation}
\int d^4x \int d^4\theta\, \lambda U .
\end{equation}
Referring to these theories by their global symmetry groups, we see
that only the global $SO(8)$ theory has an exactly marginal coupling.
This coupling is in fact the gauge coupling of the scale invariant
$SU(2)$ theory with four fundamental hypermultiplets described in
\cite{sw9408}.  The three global $U(n)$ theories were found in
\cite{apsw9511} by tuning masses and vevs in the global $SO(8)$
theory.

What interests us here are the three $E_k$ theories found in
\cite{mn96} which have no known embedding in $d=4$ gauge theories.  In
other words they have not been shown to appear as limits of other AF
or scale invariant gauge theories in four dimensions.  There is in
fact a large class of such ``exceptional'' or ``isolated'' $d=4$
conformal field theories: many can easily be found with 2 complex
dimensional Coulomb branches (though no complete classification is
known in this case), and evidence from the AdS/CFT correspondence
\cite{agmoo9905} shows that the global $E_k$ theories have relatively benign
large $N$ limits (where $N$ refers to the ``rank'' of the theory, in
this case given by the dimension of its Coulomb branch).

Given that these theories have no known embedding in a $d=4$ theory
with a weakly coupled limit, how do we know they actually exist at all
as consistent field theories?  The evidence for their existence comes
from string constructions: they can be realized as compactifications
of similar isolated $d=5$ \cite{s9608,g9608} or $d=6$ conformal
theories \cite{gh9602,sw9603,s9609} which in turn have known string
constructions.  For instance, the $d=5$ theories can be constructed
\cite{s9608} as the field theory on coincident D4 brane probes of a
IIA string background with D8 branes and O8 orientifold planes.

The question naturally arises as to what is an intrinsically four
dimensional {\em definition} of these theories.  Possible ways of
answering this question would be to either embed these theories in
some higher rank AF theories (possibly with less supersymmetry); or to
find something analogous to the lattice definition of AF gauge
theories, {\em i.e.}\ a direct nonperturbative definition.

This question is equivalent to asking for a $d=4$ characterization of
the universality classes of these theories.  Note that these isolated
$d=4$ conformal field theories should be contrasted to other $d=4$
theories which are thought to have only higher-dimensional
definitions, such as the ``scale invariant'' $d=4$ $N=2$ theories with
compact Coulomb branch \cite{i9909}, or the infrared (IR) free $d=5$
$N=2$ superYang-Mills theory \cite{r9702}.  Unlike the $d=4$ conformal
field theories we have been discussing, there is no reason to think
that these latter theories {\em have} to have a $d=4$ (or $d=5$)
definition, since they include irrelevant operators which may indeed
be inherited from a higher dimensional theory.

\subsection{Little string theories}

Perhaps the strangest and most interesting strong coupling behavior
yet encountered in supersymmetric field theories are the ``little
string'' theories which can be thought of either as $d=6$
non-gravitational string theories at a fixed (strong) coupling, or as
quasi-local field theories which lack operators more localized than a
certain length scale.

These theories are known through their string constructions.  Below I
will simply list some of the properties of these theories in the
simplest case of $d=6$ with 16 supercharges.  Their string
construction is as the theory living on a stack of $k$ NS5 branes in
either IIA or IIB string theory with fixed string tension
$1/\alpha'=M_s^2$ and in the limit as the string coupling vanishes
$g_s\to0$.  This limit implies that bulk ($d=10$) string modes
including gravity decouple, and leaves a non-trivial $d=6$ theory with
a scale $M_s$ \cite{brs9704,s9705}.

The IR properties of these theories, {\em i.e.}\ at scales below $M_s$,
are as follows (short reviews are \cite{i9708,a9911}).  At the origin of
their moduli spaces one finds either an $N=(1,1)$ supersymmetric IR free
Yang-Mills theory with a general rank $k$ gauge group $G_k$ \cite{w9710}
(from the IIB string construction); or an $N=(2,0)$ conformal field
theory with a $G_k = A_k$-$D_k$-$E_k$ classification (from the IIA
string construction).  At a generic point on moduli space the $N=(1,1)$
theory is broken down to $k$ massless $U(1)$ vector multiplets, while
the $N=(2,0)$ theory is broken down to $k$ self-dual antisymmetric
tensor multiplets.  These theories both have BPS string states with
tensions which approach $M_s^2$ at the origin in the $N=(1,1)$ theory
and zero in the $(2,0)$ theory.  The density of states at the origin
depends on the  rank as $k^2$ in the $(1,1)$ theory and $k^3$ in the
$(2,0)$ theory.  The long wavelength limit of compactifications of
both these theories on a circle gives a $d=5$ $N=2$ superYang-Mills theory
with gauge group $G_k$.

The $(2,0)$ low energy properties strongly hint at some kind of
``nonabelian self-dual 2 form gauge theory''.  However no such local
classical field theory has been constructed, despite interesting
guesses at the algebraic structure which might play a role analogous
to the Lie algebra structure of Yang-Mills theory (for example,
\cite{almy9906}), and systematic computations of the self-interactions
of 2 form gauge fields from the effective action of 5 branes (for
example, \cite{hsw9702}).  In any case, the IR properties of the
little string theories by themselves are consistent with them being
some relevant deformation of a $d=6$ local superconformal field
theory.  The real surprise is that this is {\em not} the case: the
theory is in fact nonlocal above the scale $M_s$.

The most striking evidence of this is the fact that these theories
inherit the T-duality of their IIA and IIB superstring parents.  Thus,
the $(2,0)$ theory compactified on a circle of radius $R$ is
equivalent to the $(1,1)$ theory compactified on a circle of radius
$(M_s^2 R)^{-1}$.  This sort of behavior is characteristic of closed
string degrees of freedom with tension $M_s^2$.  Furthermore, these
theories have a Hagedorn density of states $\rho(E) \sim e^{E/T_H}$
with $T_H \sim M_s/\sqrt{k}$ \cite{abks9808}.  These high energy
properties suggest that there may be a weakly coupled (and
non-gravitational) description of these theories as strings, though
perhaps only at high enough energy densities.

The question here, as in Section~2.2, is what is an intrinsic or
minimal definition of little string theory.  Actually, there is
already an answer: the discrete light cone quantization of little
strings in (uncompactified) $d=6$ is equivalent to the $N\to\infty$
limit of a certain $SU(N)$ supersymmetric quantum mechanics or a $1+1$
dimensional sigma model \cite{abkss9707,w9707,gs9712}.  Although in
principle this gives a nonperturbative definition of the theory, it is
very hard to compute with it.  Moreover, basic properties of the
theory, such as Lorentz invariance and its BPS spectrum are not
manifest and are only expected to be recovered in the $N\to\infty$
limit.  On the other hand, it has been argued that these theories obey
the axioms of ``quasi-local'' field theory in which correlators exist
in momentum space but not in position space on length scales less than
$M_s^{-1}$ \cite{pp9809,ab9812,k9912}.  Perhaps this indicates that
there is a more direct definition of little string theories.

\end{document}